\documentclass[useAMS,usenatbib,usegraphicx]{mn2e}

\title[Time-dependent simulations of steady C-type shocks]
{Time-dependent simulations of steady C-type shocks}
\author[S. Van Loo et al. ]
{S. Van ~Loo,$^{1}$\thanks{E-mail: physvl@leeds.ac.uk}, 
I.~Ashmore$^1$, P.~Caselli$^1$, S.~A.~E.~G.~Falle$^2$, and T.~W.~Hartquist$^1$ \\
$^1$ School of Physics and Astronomy, University of Leeds, Leeds LS2 9JT, UK\\
$^2$ Department of Applied Mathematics, University of Leeds, Leeds LS2 9JT, UK}
\begin{document}

\date{Accepted - . Received - ; in original form -}

\pagerange{\pageref{firstpage}--\pageref{lastpage}} \pubyear{}

\maketitle

\label{firstpage}

\begin{abstract}
Using a time-dependent multifluid, magnetohydrodynamic code, we calculated the 
structure of steady perpendicular and oblique C-type shocks in dusty plasmas. 
We included relevant 
processes to describe mass transfer between the different fluids, radiative 
cooling by emission lines and grain charging and studied the effect of 
single-sized and multiple sized grains on the shock structure. 
Our models are the first of oblique fast-mode molecular shocks in which 
such a rigorous treatment of the dust grain dynamics has been combined with
a self-consistent calculation of the thermal and ionisation structures
including appropriate microphysics.
At low densities the grains do not play any significant r\^ole in the shock 
dynamics. At high densities, the ionisation fraction is sufficiently low 
that dust grains are important charge and current carriers and, thus,
determine the shock structure. We find that the magnetic field in the shock
front has a significant rotation out of the initial upstream plane. This is 
most pronounced for single-sized grains and small angles of the shock normal 
with the magnetic field. Our results are similar to previous studies 
of steady C-type shocks showing that our method is efficient, rigorous and 
robust. 
Unlike the method employed in the previous most detailed treatment of dust
in steady oblique fast-mode shocks, ours allows a reliable calculation even 
when chemical or other conditions deviate from local statistical equilibrium.
We are also able to model transient phenomena.
\end{abstract}


\section{Introduction}
Molecular outflows 
associated with proto-stellar objects \citep[e.g.][]{BL83} drive 
shocks into dark molecular clouds where the fractional ionisation $\chi$ 
is low ($\chi < 10^{-6}$). The low values of $\chi$ cause the 
gas and the magnetic field to be weakly coupled.
 Charged particles are then pushed through the neutral gas by 
Lorentz forces, giving rise to ambipolar diffusion \citep{MS56}. 
Collisions between the charged and neutral particles
accelerate, compress and heat the neutral gas ahead of the shock front 
and a magnetic precursor forms. If the shock speed is low enough for 
the gas to cool efficiently, the flow becomes continuous in 
all fluids \citep[e.g.][]{M71}. Such shocks are referred to as C-type 
shocks \citep{D80}.  

\citet[][hereafter DRD]{DRD83} calculated the structures of 
steady perpendicular C-type shocks in diffuse and dense molecular media 
using a steady-state multifluid magnetohydrodynamic (MHD) description. 
They included a detailed treatment for energy and momentum transfer 
between different particles, oxygen-based chemistry and radiative 
cooling. They also assumed that the ion and electron densities exceed the 
grain charge density. However, for a fractional ionisation below 10$^{-7}$,
grains play an important r\^ole in governing the electromagnetic
forces \citep{D80}. In the densest regions of molecular clouds where 
$n_H \ge 10^6~{\rm cm^{-3}}$, the fractional ionisation constraint is 
usually fulfilled.

\citet{PHH90} included the effect of dust grains 
on the structure of perpendicular steady C-type shocks 
more rigorously than previous authors (e.g. DRD).
Their four-fluid models predict 
different shock widths than obtained in similar models of DRD. 
This model was extended to oblique shocks by
\citet{PH94} who found that in  
such shocks the tranverse component of the magnetic field 
rotates about the propagation direction of the shock as long as there 
is a non-negligible Hall conductivity. This happens for a wide range of 
conditions, but Pilipp \& Hartquist were unable to find solutions for 
fast-mode shocks. 

\citet{W98} pointed out that, to find a stable shock structure for steady 
oblique fast-mode shocks, one needs to integrate the steady state equations 
backwards in time from the downstream boundary. However, such a method 
requires an exact thermal, chemical and ionisation equilibrium. 
Such an equilibrium does not obtain in most shocks of interest. For
instance, the abundance of water, an important coolant, is out of equilibrium
for about 10$^5$ years or more from the time the gas cools below several
hundred degree. This highlights the necessity for time-dependent 
simulations to calculate the C-type shock structure.  
\citet{CR02} were the first to include a fluid description of 
grain dynamics in multifluid MHD simulations. However, their study is 
restricted to perpendicular shocks. 
The multifluid approach developed by \citet{F03} overcomes this limitation.  

In this paper we study the structures of steady fast-mode 
C-type shocks using time-dependent multifluid MHD simulations. In 
Sect.~\ref{sect:model} we describe the numerical code and the physical 
processes that we include. Then, we apply the code to perpendicular and 
oblique shocks with parameters similar to those of DRD including 
a single-sized grain fluid (Sect.~\ref{sect:single}) or multiple grain 
fluids (Sect.~\ref{sect:multiple}). Finally, we discuss the 
results and give our conclusions in Sect.~\ref{sect:conclusions}.

\section{The model}\label{sect:model}
\subsection{Numerical code}
As the ionisation fraction within molecular clouds is low, the plasma
needs to be treated as a multicomponent fluid consisting of neutrals,
ions, electrons and $N$ dust grain species. The governing equations for 
the neutral fluid are given by
\begin{eqnarray}\label{eq:neutral} 
    \frac{\partial{\rho_n}}{\partial{t}} + \frac{\partial}{\partial{x}}
	(\rho_n v_{x,n})  &=&  \sum_{j} s_{jn}, \label{eq:rhon} \\
    \frac{\partial{\rho_n {\bf v}_n}}{\partial{t}} + 
	\frac{\partial}{\partial{x}}(\rho_n v_{x,n}
        {\bf v}_n + p_n {\bf 1_{x}})  &=&  \frac{{\bf J}\times{\bf B}}{c}, \label{eq:monentumn}\\
  \frac{\partial{e_n}}{\partial{t}} + \frac{\partial}{\partial{x}} 
	\left[v_{x,n} \left(e_n + p_n + \frac{1}{2} \rho_n v_n^2 \right) 
	\right]  &=& {\bf J}.{\bf E} + \sum_{j} H_j, \label{eq:energyn}
\end{eqnarray}
where $e_n = p_n/(\gamma - 1) + \rho_n v_n^2/2$ is the internal energy per unit 
mass, $s_{jn}$ the mass transfer rate between the charged fluid $j$ and the 
neutral fluid, $H_j$ the energy sinks or sources for the different fluids, 
$\bf{B}$ the magnetic field, $\bf{E}$ the electric field and
$\bf{J}$ the current given by $\bf{J} = \sum_j{\alpha_j \rho_j \bf{v}_j}$.

In the limit of small mass densities for the charged fluids, their inertia 
can be neglected. 
Also, the charged fluids are in thermal balance as their heat capacities
are small.
For the charged fluid $j$, the equations then reduce to
\begin{eqnarray}
     \frac{\partial{\rho_{j}}}{\partial{t}} + \frac{\partial}{\partial{x}}
	(\rho_{j} v_{x,j})  &=&  -s_{jn}, \label{eq:rhoc}\\    
     \alpha_{j} \rho_{j} \left({\bf E} + \frac{{\bf v}_j\times {\bf B}}{c}
	\right) + 
        \rho_{j} \rho_n K_{jn} ({\bf v}_n - {\bf v}_j)  &=&  0, \label{eq:momentumc}\\
     H_{j} + G_{jn}  &=& 0, \label{eq:energyc}
\end{eqnarray}
where $\alpha_j$ is the charge to mass ratio, $K_{jn}$ the collision rate
coefficient of the charged fluid $j$ with the neutrals and $G_{jn}$ the 
energy transfer rate per unit volume between the charged fluid and the 
neutral fluid. 
The expressions for $K_{jn}$ and $G_{jn}$ are given by \citet{D86}.
We use the reduced energy equation to calculate the temperatures 
of the ion and electron fluids. For the grain fluids, a hydrodynamic 
approximation with zero pressure is appropriate as the thermal velocity 
dispersion of the dust grains is small compared to the drift velocity with 
the neutrals.

The collisions of the charged particles with the neutral particles 
affect the evolution of the magnetic field. Using the reduced momentum
equations for the charged fluids (Eq.~\ref{eq:momentumc}) and the expression
for the current $\bf{J}$, the electric field can be expressed 
as \citep{C56} 
\[
   {\bf E} = -{\bf v_n} \times {\bf B} + r_{//} 
	\frac{({\bf J}.\bf{B})\bf{B}}{B^2} + r_H \frac{{\bf J} \times \bf{B}}{B}
	- r_{AD} \frac{({\bf J} \times \bf{B}) \times \bf{B}}{B^2}, 
\] 
where $r_{//}$ is the resistivity along the magnetic field, $r_H$ the 
Hall resistivity and $r_{AD}$ the ambipolar resistivity. 
The expressions of the resistivities are given by e.g. \citet{F03} and 
depend strongly on the Hall parameter, 
$\beta_i = \alpha_i|{\bf B}|/K_{ni} \rho_n$, which is the ratio 
between the gyrofrequency of the charged fluid and collision frequency of 
the charged fluid with the neutrals. 
Substitution of the above expression into the Maxwell-Faraday 
equation  then gives the magnetic induction equation 
\begin{equation}\label{eq:Bfield}
\frac{\partial{{\bf B}}}{\partial{t}} + \frac{\partial{{\bf M}}}{\partial{x}}
  = \frac{\partial}{\partial{x}} {\bf R} \frac{\partial{\bf B}}{\partial{x}},
\end{equation}		
where ${\bf M} = {\bf v_n} \times {\bf B}$ is the hyperbolic magnetic flux
and ${\bf R}$ is the resistance matrix which depends on the Hall resistivity,
the ambipolar resistivity, and the resistivity along the magnetic field.
(For an expression for ${\bf R}$, see \citet{F03}.)	
  
Equations~\ref{eq:neutral} -~\ref{eq:Bfield} are solved using the numerical
scheme described by \citet{F03}.  This scheme uses a second-order 
hydrodynamic Godunov solver for the neutral fluid equations 
(Eqs.~\ref{eq:rhon}-\ref{eq:energyn}). The charged fluid densities are 
calculated using an 
explicit upwind approximation to the mass conservation equation
(Eq.~\ref{eq:rhoc}), while the velocities and temperatures are  
calculated from the reduced momentum and energy equations 
(Eq.~\ref{eq:momentumc} and \ref{eq:energyc}). The magnetic field is 
advanced explicitly, 
even though this implies a restriction on the stable time step at high 
numerical resolution \citep{F03}.

\subsection{Physical processes}
The numerical code described in the previous section is quite general. 
In order to investigate the effects of dust grains in fast-mode shocks 
it is necessary to specify the source terms, i.e. the mass, momentum and 
energy transfer coefficients and the energy sources and sinks, to 
describe the relevant physics. 
The chemical network that we adopted is the one used
by \citet{PHH90} and \citet{PH94}.

Mass transfer between the ion and electron fluids and the neutral fluid 
is due to cosmic ray ionisation at a rate of $10^{-17}~{\rm s}^{-1}$, electron 
recombination with Mg$^+$, dissociative recombination with HCO$^+$ 
and recombination of ions on charged grains.
Here we have assumed that the main gas phase ions are Mg$^+$ and  HCO$^+$ 
\citep{OD74}. The cross-section for 
these recombination reactions are taken from \citet{PHH90}.

Mass is also transferred between the ion and electron fluids and the 
grain fluids due to ion neutralisation on grain surfaces. Ions and 
electrons stick to dust grains upon colliding with them.  Also, they 
pass on their charge with unit efficiency. The grain charge then critically 
depends on the electron-grain and ion-grain collision rates (which 
themselves depends on the electron density and temperature, the ion 
density and temperature, and the ion-grain relative streaming speed). 
The average grain charge is evaluated following \citet{HHP87}. Individual
grains can have a charge that differs from the average value, but it is  
mostly within one unit of the average \citep{E79}.

Different radiative cooling processes for the neutral and electron fluids
are included in the code. The electron gas is cooled by the excitation and 
subsequent de-excitation of molecular hydrogen (e.g. DRD). This process 
only becomes important when the electron fluid temperature exceeds 
$\approx 1000\ K$. In the neutral gas, the dominant cooling processes 
are associated with O, CO, H$_2$ and H$_2$O. 
The radiative transitions for 
atomic oxygen, i.e. OI lines, are calculated using the procedure of DRD. 
The radiative losses from excited rotational-vibrational states of molecular 
hydrogen are a fit to the data presented by \citet{HDO80}, while those of CO 
are calculated by using \citet{HM79}. Finally, we use \citet{NM87} to 
calculate the losses due to rotational transitions in H$_2$O. The heating 
rate of the neutral gas by cosmic rays is from DRD.  
We have not included any radiative losses for the ion fluid. 
From Eq.~\ref{eq:energyc} and using the expression for $G_i$ \citep{D86}
we find that the ion temperature, $T_i$, is given by 
\begin{equation}\label{eq:Ti}
    T_i = T_n + \frac{m_n}{3k_B} ({\bf v}_n - {\bf v}_i)^2,
\end{equation}
where $T_n$ is the neutral temperature, $m_n$ the neutral particle mass
and $k_B$ the Boltzmann constant. 

\begin{figure*}
\includegraphics[width = 6.4 cm]{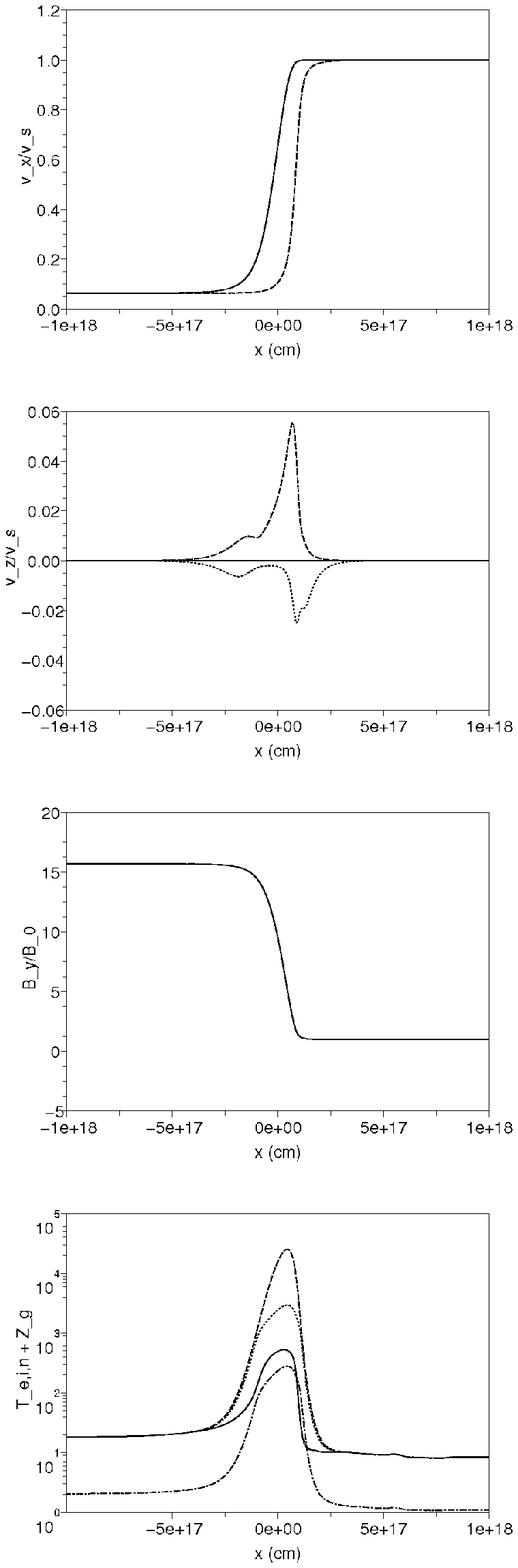}
\includegraphics[width = 6.4 cm]{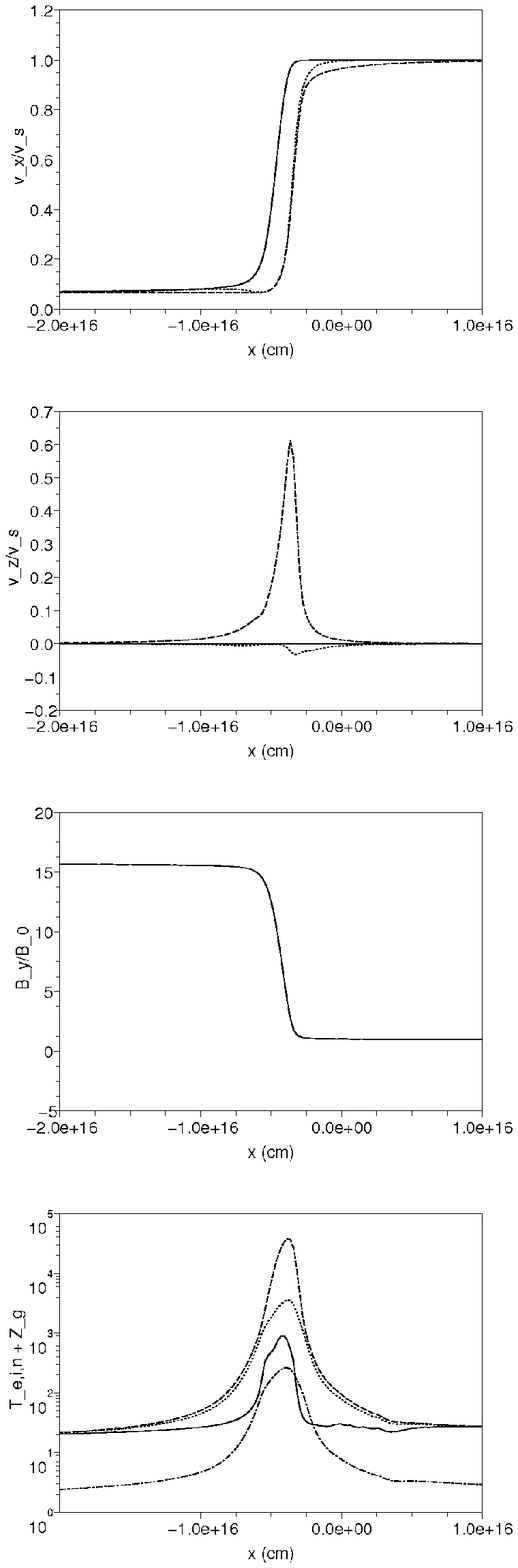}
\caption{The shock structure for a perpendicular shock propagating in a 
quiescent region with $n_{\rm H} = 10^4 {\rm cm^{-3}}$ (left) and 
$n_{\rm H} = 10^6 {\rm cm^{-3}}$ (right). From top to bottom we plot 
the velocity
in the $x$-direction and $z$-direction for the neutrals (solid), 
ions/electrons (dashed) and grains (dotted), the tangential magnetic 
field and the temperatures of the ions (dashed), electrons 
(dotted) and the neutrals (solid) as well as the absolute value of the 
grain charge (dash-dotted). The velocities are normalised with respect to 
the shock speed and the magnetic field to the total upstream magnetic field.}
\label{fig:perp}
\end{figure*}

\subsection{Initial conditions}\label{sect:initial conditions} 
The initial conditions of our simulations are adopted from DRD and 
similar studies such as \citet{W98} and \citet{CW06}. This allows a  
comparison of our results with those of these studies. 

The neutral fluid has a number density of $n_{\rm H} = 10^4\ {\rm cm^{-3}}$ or 
$n_{\rm H} = 10^6\ {\rm cm^{-3}}$ and is composed mainly of molecular hydrogen 
so that $m_n \approx 2 m_{\rm H}$. Small amounts of other neutral species such 
as O, CO, and H$_2$O are also present. The initial abundances of O, CO, and 
H$_2$O are taken to be 4.25$\times 10^{-4}~n_{\rm H}$, 
5 $\times$ 10$^{-5} n_{\rm H}$ and 0 respectively. However, as the abundances 
of 
O and H$_2$O evolve substantially as gas is heated, the abundances 
of the important coolants
are recalculated at  every time-step. 
The dominant ions within the gas are assumed, as mentioned above, 
 to be HCO$^{+}$ and Mg$^+$. The ion mass is then $m_i \approx 30 m_{\rm H}$. 
Also, we assume that the ions are only singly ionised. 

The dust grains are all assumed to be spherical with a radius of 
$a_g = 0.4 \mu m$ and a mass of 8.04 $\times 10^{-13}~g$ (where the density 
of the grains is taken 3 g\ cm$^{-3}$). The grain to hydrogen mass ratio
in the unshocked region has the canonical value of 0.01. 
The densities of the ion and electron fluids and the average grain charge
are calculated from ionisation equilibrium. For the temperatures that we
adopt, i.e. $T_n = 8.4K$ for $n_{\rm H} = 10^4\ {\rm cm^{-3}}$ and 
$T_n = 26.7K$ for $n_{\rm H} = 10^6\ {\rm cm^{-3}}$, the average grain 
charge is quite close to 
$-e$ and the electron and ion number densities are roughly 
the same, i.e. $n_e \approx n_i$. 

In the interstellar medium the magnitude of the magnetic field scales roughly
as $B \approx 1 \mu G (n_{\rm H}/{\rm cm^{-3}})^{-1/2}$ (see DRD) so that 
B = 10$^{-4}$ G for $n_{\rm H} = 10^4 {\rm cm^{-3}}$ and B = 10$^{-3}$ G for 
$n_{\rm H} = 10^6 {\rm cm^{-3}}$. The Alfv\'en speed of the gas 
is then the same at both densities, i.e. $v_A = 2.2~{\rm km\ s^{-1}}$. In the 
far upstream region the magnetic field 
lies in the $x,y$-plane and makes an angle of 30$^\circ$, 45$^\circ$, 
60$^\circ$ or 90$^\circ$ (perpendicular shock) with the $x$-axis. We assume 
that a shock is propagating in the positive $x$-direction with 
a speed of 25~km\ s$^{-1}$.  Such a shock is a strong fast-mode shock 
with an Alfv\'enic Mach number of $M_A \approx 11.5$.

Using the upstream fluid properties, the properties far downstream 
of the shock are calculated from the Rankine-Hugionot relations for an 
isothermal, magnetised shock. For our initial shock structure we assume that 
the shock is a discontinuity at $x=0$ separating the upstream and downstream 
properties. We follow the evolution of the shock to steady state within 
the shock frame. 

\subsection{Computational details}\label{sect:computational details}
The size of the computational box is assumed to be a few time the shock 
width.  By balancing the ion-neutral drag 
force with the Lorentz force, we can estimate the shock thickness 
\citep[e.g.][]{DM93}
\begin{equation}\label{eq:ionwidth}
	L_i = 7\times 10^{15} \left(\frac{v_A}{{\rm km\ s^{-1}}}\right) \left(
	\frac{10^{-2} {\rm cm}^{-3}}{n_i} \right) {\rm cm},
\end{equation}
where $v_s$ is the shock speed.  However, the grain-neutral friction 
is comparable to the ion-neutral friction for 
\[
 \frac{\rho_i}{\rho_g} < \frac{K_{gn}}{K_{in}},
\]
or 
\[
   \chi < 3 \times 10^{-9} \left(\frac{v_s}{\rm km\ s^{-1}}\right),
\]
where we 
substituted the expressions for the collision rates 
with the neutrals \citep{D86} and
assumed the relative speed between the neutrals and the grains
to be of the order $v_s/2$.  This constraint is fulfilled 
in the heated precursor for both densities and the shock thickness
is determined by the grain-neutral drag. The expression for the 
shock width is similar to Eq.~\ref{eq:ionwidth} and is given by 
\begin{equation}\label{eq:width}
     L_g = 3 \times 10^8 \chi \left(\frac{v_s}{\rm km\ s^{-1}}\right)^{-1} L_i.
\end{equation}
Values of $\chi$ appropriate for the precursor must be used. \citet{FPH85}
first pointed out that chemistry leads to a large drop in $\chi$ in the 
precursor of many shocks.  The numerical domain is then chosen to 
be  $-10L_g \leq x \leq 10 L_g$ with a free-flow boundary condition at the 
downstream side and fixed inflow at the upstream boundary. We use a uniform 
grid with 400 cells. 

As our initial conditions are discontinuous, the inertia of the 
charged fluid, even the ions, cannot be neglected during the initial stages 
of the evolution. However, the inertial phase of the charged fluid is 
short compared to the time to approach a steady configuration 
\citep[e.g.][]{CR07}. Although the inertial phase does not affect the final 
steady shock structure here, this effect should not be ignored for some other 
time-dependent models.

\begin{figure*}
\includegraphics[width = 5.6 cm]{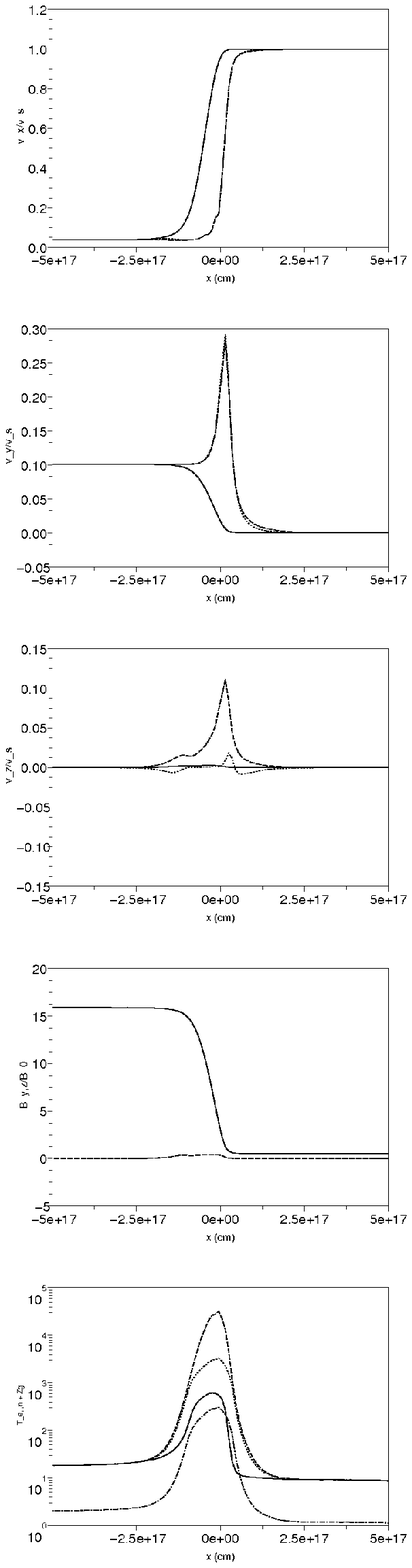}
\includegraphics[width = 5.6 cm]{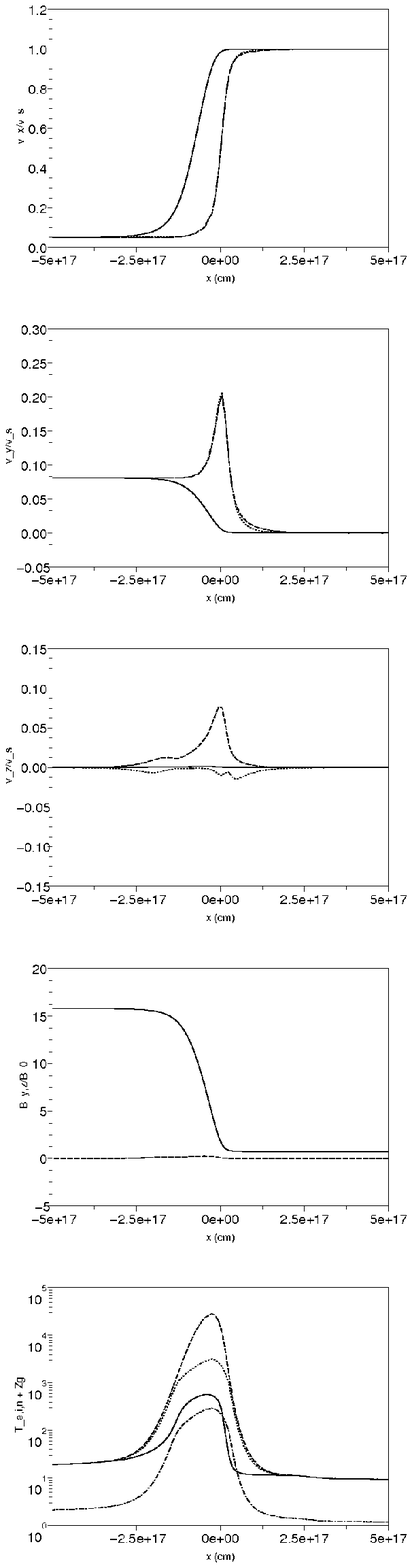}
\includegraphics[width = 5.6 cm]{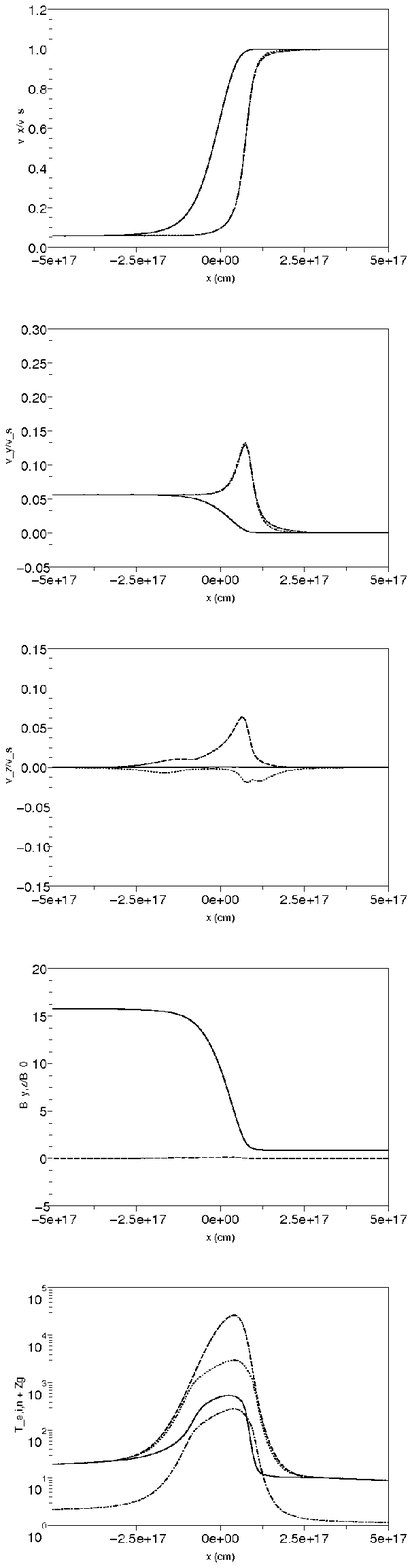}
\caption{The shock structure for an oblique shock 
propagating at an angle of 30$^\circ$ (left), 45$^\circ$ (middle) and 
60$^\circ$ (right) with the magnetic field in a quiescent region with 
$n_{\rm H} = 10^4 {\rm cm^{-3}}$. From top to bottom we plot the velocity
in the $x$-direction, $y$-direction and  $z$-direction for the neutrals (solid),
ions/electrons (dashed) and grains (dotted), the tangential magnetic
field in the $y$-direction (solid) and $z$-direction (dashed), and the 
temperatures of the ions (dashed), electrons (dotted) and the 
neutrals (solid) as well as the absolute value of the
grain charge (dash-dotted). The velocities are normalised with respect to
the shock speed and the magnetic field to the total upstream magnetic field.}
\label{fig:oblique4}
\end{figure*}

\section{C-type shock models}\label{sect:single}
In this section we present steady C-type shock structures for 
large single-sized grains. First we discuss perpendicular shocks 
and then shocks propagating at an oblique angle with the magnetic field.   

\subsection{Perpendicular shocks}
Figure~\ref{fig:perp} shows the structures in multiple flow variables 
for perpendicular shocks at different densities. It is clear that both 
shock structures represent 
C-type shocks. As our models are similar to DRD, we can compare our results
with figs.~2 and 3 of DRD. Although the shapes of the shocks are 
qualitatively similar, our shock widths are about 4 times larger than 
in DRD. This is due to a lower fractional ionisation in our models which 
is a consequence of including the chemistry governing the ionisation rather
than assuming a constant ion flux as DRD did. 
The drag force of the charged particles on the neutrals is then lower 
as well and is balanced by the magnetic pressure over a larger length scale.
A consequence of the 
larger shock widths is that the gas has more time to radiate away its
energy. The maximum temperature of the neutrals within the shock front 
is only 530 K for $n_{\rm H} = 10^4 {\rm cm^{-3}}$ and 905 K for 
$n_{\rm H} = 10^6 {\rm cm^{-3}}$, while DRD found 1223 K and 1334 K 
respectively.

Figure~\ref{fig:perp} shows that the speed of the grains relative to the 
neutrals is less than the speed of the ions relative to the neutrals.
The difference in relative speeds is due to the different response of 
the charged fluids to the Pedersen along $\bf{E'}$ ($= \bf{E} + \bf{v}_n \times
{\bf{B}}$) and Hall current along $\bf{E'}\times\bf{B}$ within the shock front.
From Eq.~\ref{eq:momentumc}  we find 
\begin{equation}\label{eq:velocity}
  \frac{\bf{v_j} - \bf{v_n}}{c} \approx \frac{\beta_j^2}{1+\beta^2_j}
  \left(\frac{\bf{E'} \times \bf{B}}{B^2}\right) +  \frac{\beta_j}{1+\beta^2_j}
   \left(\frac{\bf{E'}}{B}\right),
\end{equation}
where $\beta_j$ is the Hall parameter of the charged fluid. For electrons
and ions with $|\beta_{e,i}| \gg 1$, the particles are tied to the
magnetic field and 
\[
\frac{\bf{v_{e,i}} - \bf{v_n}}{c} \approx \frac{\bf{E'} \times \bf{B}}{B^2}.
\]
At the other extreme, $|\beta_j| \ll 1$, the charged particles move with the 
neutrals. For the $n_{\rm H} = 10^6 {\rm cm^{-3}}$ models, the grains have 
a Hall parameter ranging from -0.1 (the upstream value) to a minimum of -1.5 
within the shock front. For these values, the $\bf{E'}$ term has a 
non-negligible contribution to the drift velocity. For the $x$-component 
of the velocity of the grains, this term only becomes important when 
the grains switch from moving with the neutrals to moving with the 
ions and electrons (see Fig.~\ref{fig:perp}). The $z$-component of the 
$\bf{E'}$ term, however, has a larger contribution to the drift forcing 
the grains to move with the neutrals.
The lower density model exhibits this effect to a lesser extent as the 
grain Hall parameter lies between -1 and -20. Then the  $\bf{E'}$ term is 
less important and provides only a small contribution to the drift. 

The velocity difference between the ions and electrons with the neutrals 
determines the temperature of the charged particles through collisional 
heating. The maximum velocity difference in the models is $\approx v_s$ 
giving a maximum ion temperature of a few times
10$^4$ K. Electron cooling becomes important at temperatures
of about 1000\ K reflected in a considerably lower electron temperature 
at roughly 3$\times 10^3$~K. The electron temperature within the shock 
front is important for the dynamics of the shock as the average grain charge 
is 
\begin{equation}\label{eq:z}
	Z_g e \approx -\frac{4 k_b T_e a_g}{e},
\end{equation}
for $|Z_g| \gg 1$ and $n_g|Z_g| \ll n_e$ \citep[e.g.][]{D80}. 
Otherwise, the average grain charge is roughly $-1$. From 
Fig.~\ref{fig:perp} we find that the average grain charge follows
this dependence on the electron temperature with the average grain charge 
reaching maximum values of about -250. 

\begin{figure*}
\includegraphics[width = 5.6 cm]{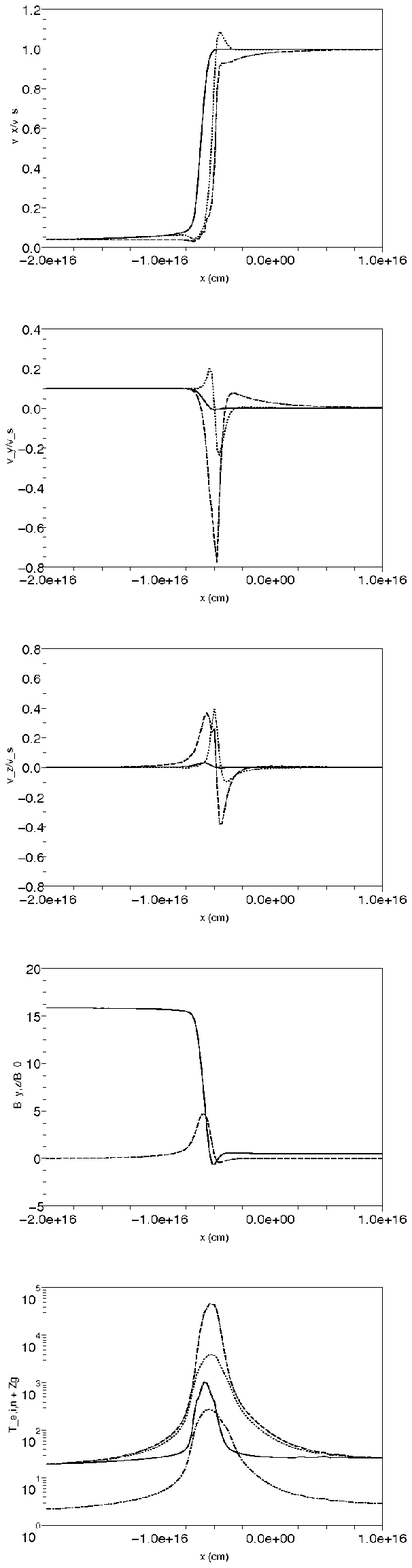}
\includegraphics[width = 5.6 cm]{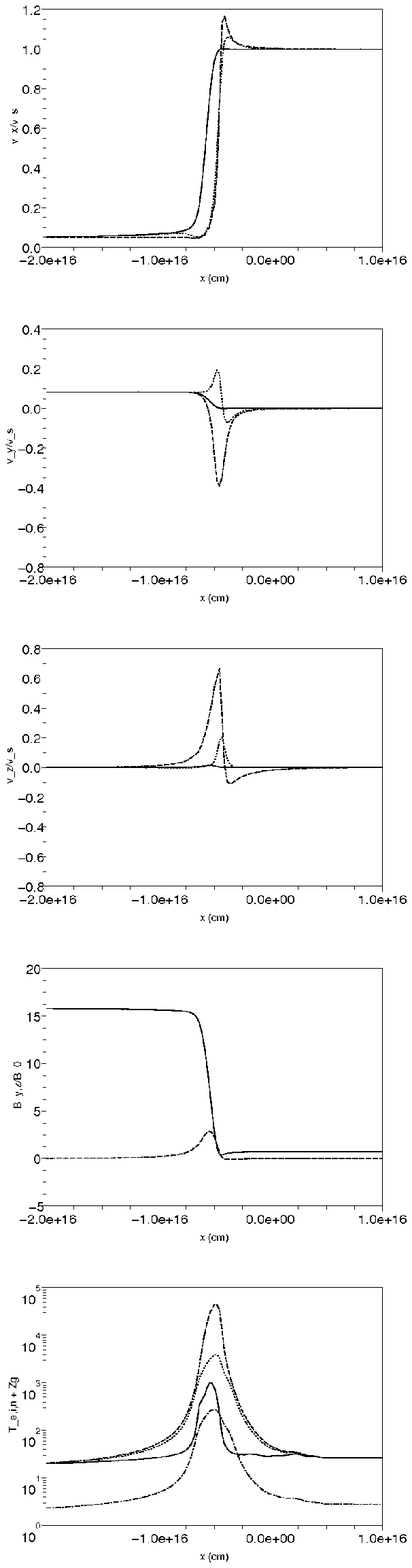}
\includegraphics[width = 5.6 cm]{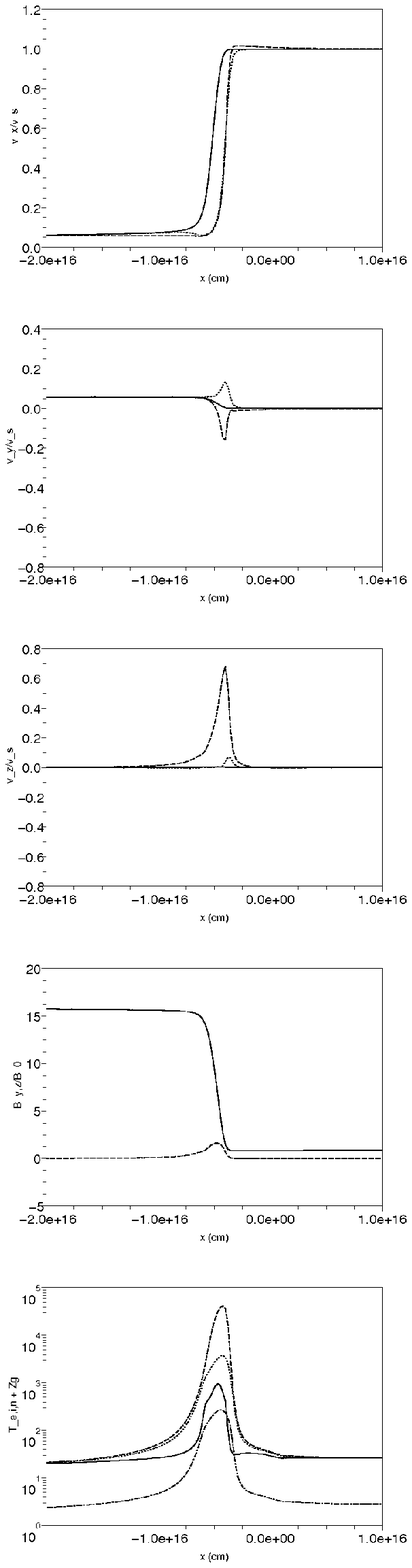}
\caption{Same as in Fig.~\ref{fig:oblique4}, but with an upstream density of
$n_{\rm H} = 10^6 {\rm cm^{-3}}$.}
\label{fig:oblique6}
\end{figure*}

\subsection{Oblique shocks}
For a finite magnetic field component parallel to the shock propagation
and a non-zero Hall conductivity, the Hall current along $\bf{E'}\times\bf{B}$ 
has a  component parallel to the propagation direction of the shock. As the 
total current in that direction is zero if the shock is steady, 
the Hall current needs to be 
balanced by the Pedersen current (along $\bf{E'}$). This Pedersen current 
gives rise to the rotation of the magnetic field around the shock 
propagation \citep{PH94}. Significant rotation occurs when the Hall 
conductivity is larger than the Pedersen conductivity \citep{W98}.  
  
For the $n_{\rm H} = 10^4\ {\rm cm^{-3}}$ models, the Hall conductivity is 
always smaller than the Pedersen conductivity and there is no significant  
rotation of the magnetic field (see $z$-component of magnetic field 
in Fig.~\ref{fig:oblique4}).
The rotation increases marginally as the angle, $\theta$, between 
the direction of shock propagation and the magnetic field decreases.
The other shock properties, also, do not change significantly with 
$\theta$. The temperatures of the different fluids and the average 
grain charge are similar to the ones found for the perpendicular shock.
These results agree well with the $\theta$-dependence of the steady 
shock structure found by \citet{CW06}. The similarity of the shock
structure for different values of $\theta$ results from the minor contribution 
of the grains to the drag force on the neutrals. The grains do not 
determine the dynamics and drift with the other charged species. 

Significant differences, however, are seen in the 
$n_{\rm H} = 10^6~{\rm cm^{-3}}$
models. Here, the grains are the dominant contributors to the drag force on
the neutrals. Furthermore, there is a significant charge separation giving 
rise to a large Hall conductivity. In Fig.~\ref{fig:oblique6} we 
see that there is a substantial rotation of the magnetic field. Also the 
magnetic field upstream is no longer a stable node, but a spiral node as
the Hall conductivity changes wave propagation upstream \citep{W98, F03}.
As for $n_{\rm H} = 10^4 {\rm cm^{-3}}$, the rotation decreases  for 
larger $\theta$, while the shock width increases.

The rotation of the magnetic field has large consequences for the 
drift velocity of the charged particles. Figure~\ref{fig:oblique6} shows 
that the velocity structure changes considerably with $\theta$. Also,
the grains move significantly differently than  the ions and electrons, 
especially in the $\theta = 30^\circ$ model (see Fig.~\ref{fig:oblique6}). 
Along the propagation direction, electrons and ions are decelerated 
immediately. The grains, however, are initially accelerated
before being decelerated. Furthermore, they lag the ions and electrons.
To drift in the direction opposite to the ions and electrons, the grains
motion must be dominated by the $\bf{E'}$ term in Eq.~\ref{eq:velocity}
and $E'_x$ must be in the same direction as the $x$-component of 
$\bf{E'}\times\bf{B}$. The large differences in drift velocity are expected
as the rotation of the magnetic field is induced by a charge separation
(and thus large Hall conductivity). For larger $\theta$, the Hall conductivity
and rotation decrease, so that the ions, electrons, and grains move more
together. As expected, the velocity profile converges to the perpendicular 
shock structure (see Fig.~\ref{fig:perp} and \ref{fig:oblique6}).

\begin{figure}
\begin{center}
\includegraphics[width = 6.0 cm]{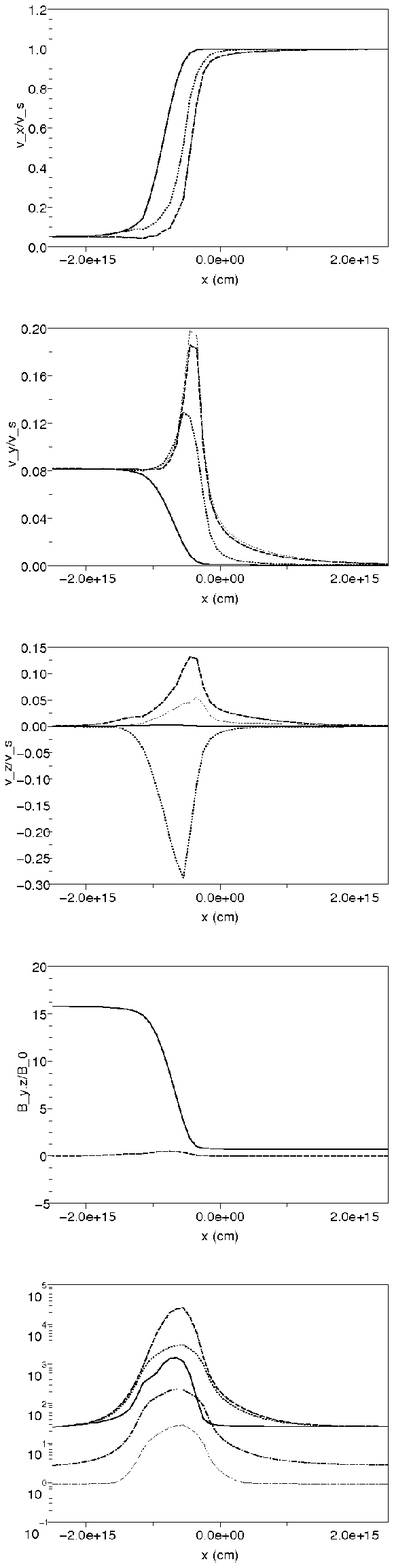}
\end{center}
\caption{Similar to the $45^\circ$ model of Fig.~\ref{fig:oblique6},
but for two grain fluids. The small grains are indicated 
on the velocity components figures with a thin-dotted line, while their average
charge is given by a thin dashed-dotted line. }
\label{fig:oblique5f}
\end{figure}

\section{Multiple grain species}\label{sect:multiple}
In Sect.~\ref{sect:single} we have assumed a single-sized grain fluid.
In molecular clouds, however, the dust has a grain-size distribution 
given by \citep{MRN77}
\begin{equation}\label{eq:MRN}
	\frac{dn}{da} \propto a^{-3.5},
\end{equation}
where $dn$ is the number density of grains with radii between $a$ and
$a + da$.
To take into account the effect of multiple dust species, we 
include an additional small-grain fluid with $a_s = 0.04 \mu m$ 
(i.e. $a_s = a_g/10$) and $m_s = 8.03\times 10^{-16}$g (or $m_s = m_g/1000$). 
The total mass density of the grain fluid, $\rho_s + \rho_g$, is assumed 
to be 1\% of the neutral mass density (as in our previous models). 
With the adopted grain-size distribution most of the mass is in the 
small grain fluid. Furthermore, the small grains dominate the 
grain-neutral friction and, thus, the shock dynamics as the  
collision coefficient between grains and neutrals is proportional to 
$a_{g,s}^2/m_{g,s}$ \citep[see][]{D86}.
From Fig.~\ref{fig:oblique5f}, we see that the shock width is indeed, as 
expected from Eq.~\ref{eq:width}, an order of magnitude smaller than 
for the model with single-sized large grains. 

Although the small grains have an average charge about an order of
magnitude smaller than the large grains (see Eq.~\ref{eq:z}), the small 
grains have a larger 
Hall parameter and are strongly coupled to the magnetic field. 
The velocity plots in Fig.~\ref{fig:oblique5f} show that the 
small grains move with the electrons and ions, while the large grains 
move with velocities between those of
the neutrals and the other charged particles. The 
grain charge density is dominated by the small grains and free electrons are
depleted from the gas phase (i.e. $n_i \gg n_e$).
The charge separation between the charged particles is thus small and  
the Hall conductivity is too. The rotation of the magnetic field out 
of the plane is then negligible. Also, the upstream magnetic 
field is no longer a spiral node, but a stable node.

The inclusion of a second
grain species thus changes the shock structure considerably. 
Including additional grain species will further affect the shock
structure as more dust grains are loading the magnetic field lines. 
Then the signal speed at which disturbunces propagate through the charged 
fluid decreases. For small PAH abundances the signal speed is close 
to 25~km\ s$^{-1}$ \citep{FP03} leading to the possibility that the 
resulting shock is a J-type shock in which the grain inertia cannot be 
neglected.

\section{Summary and discussions}\label{sect:conclusions}
In this paper we have presented 1D time-dependent multifluid MHD simulations 
of perpendicular and oblique C-type shocks in dusty plasmas. We have included
in our numerical code 
relevant mass transfer processes, such as electron recombination with Mg$^+$ 
and dissociative recombination with  HCO$^+$, and radiative cooling 
by O, CO, H$_2$ and H$_2$O. Also, the mass and 
charge transfer from ions and electrons to the dust grains is 
taken into account to  calculate the average grain charge. 
Our models are the first of oblique fast-mode shocks in dense molecular
clouds in which a fluid treatment of grain dynamics has been combined
with a self-consistent calculation of the thermal and ionisation balances
including appropriate microphysics. We have performed 
simulations with single-sized and with two dust grains species in plasmas of 
different density. 

Dust grains do not change the shock dynamics much at 
$n_{\rm H} = 10^4 {\rm cm^{-3}}$
and the shock structures for oblique and perpendicular shocks are quite similar.
Grains, however, dominate the shock dynamics at high densities 
($n_{\rm H} = 10^6 {\rm cm^{-3}}$). For oblique shocks, the magnetic field 
in the shock front is not confined to the upstream magnetic field plane. 
The amount of rotation obtained depends on the Hall and Pedersen conductivities.
For single-sized grains the rotation is significant and increases with smaller
$\theta$ as the Hall conductivity in these models is comparable to the Pedersen 
conductivity. For some models, the magnetic field at the upstream side 
of the shock even spirals around the direction of shock propagation.
When we include an additional small grain species,
the rotation mostly disappears. The small grains which carry most of the 
charge move with the ions and electrons reducing the Hall conductivity.
Our models corroborate the main qualitative conclusions
of previous studies \citep[DRD][]{PH94,
W98, CW06, GPJ07}, showing that our numerical method is efficient, 
rigorous, and robust. 
In addition, we have succeeded in taking a significant step by developing
the first oblique shock code capable of producing reliable results 
for comparisons with observations. Previous oblique shock models were 
restricted by less rigorous treatments of either the grain dynamics or the
thermal and ionisation balances.

However, our model does have some restrictions.
Firstly, we determine the drift velocities of the grains solely by 
balancing the grain-neutral friction with the Lorentz force and neglect 
their inertia. While small dust grains can be considered inertialess, 
large grains have a drag length, i.e. the length scale for gas drag to 
decelerate the grains, that in some cases is
 comparable to the length scale for magnetic field 
compression \citep{CR02}. Therefore, large grain inertia cannot always 
be neglected. Secondly, for small grains, the average grain charge is 
close to the elementary charge.
As there is some fluctuation around the average, a fraction of the small 
grains has a zero or even positive charge. The neutral grains decouple 
from the magnetic field, while the positive grains drift in a different 
direction to the negative grains. Although this fluctuation around the average 
grain charge changes the shock structure, its effect is only minimal 
\citep{GPJ07}.

A time-dependent method is not only robust (even for non-equilibrium 
conditions) for finding steady state shock structures. It is
also vital for modelling transient phenomena. In the vicinity
of protostellar objects the abundance of SiO is significantly enhanced 
\citep[e.g.][]{MBF92, JS04}. It is thought that the SiO emission arises
from the interaction of a protostellar jet with local dense clumps
\citep{Letal98}.  Grain sputtering within C-type shocks releases the 
SiO depleted on the grains back into the gas phase \citep{CHH97, S97}. 
Numerical models using steady C-type shocks indeed show good agreement with 
recent observations of SiO line intensities in the L1157 and L1448 molecular 
outflows \citep{G08, G08b}. However, it is doubtful that shocks 
interaction with dense clumps have reached steady state. In subsequent 
papers, we will study the interaction of C-type shocks with dense clumps and 
include grain sputtering to calculate the SiO emission. 

\section*{Acknowledgements}
We  thank the anonymous referee for a report that helped to
improve the manuscript. SVL gratefully acknowledges STFC for the 
financial support.

\label{lastpage}
\end{document}